\documentclass[nofootinbib,a4paper,aps,prd,10pt,superscriptaddress,showkeys]{revtex4}

\usepackage{graphicx}
\usepackage{amsfonts}
\usepackage{amssymb} 
\usepackage{amsmath}
\usepackage{natbib}
\usepackage{float}
\usepackage{dcolumn}% Align table columns on decimal point
\usepackage{bm}% bold math
\usepackage{makecell}
\usepackage{multirow}
\usepackage{latexsym,color}
\usepackage[colorlinks=true, linkcolor=black, citecolor=blue, urlcolor=blue]{hyperref}

\newcommand{\bea}{\begin{eqnarray}}
\newcommand{\eea}{\end{eqnarray}}
\newcommand{\be}{\begin{equation}}
\newcommand{\ee}{\end{equation}}

\begin{document}

\title{Testing Scalar Field Dark Matter models in M31 galaxy through the Rotation Curve analysis}

\author{ Gulnara~\surname{Suliyeva}}
\email[]{g_suliyeva@mail.ru}
\affiliation{National Nanotechnology Laboratory of Open Type,  Almaty 050040, Kazakhstan.}
\affiliation{Al-Farabi Kazakh National University, Al-Farabi av. 71, 050040 Almaty, Kazakhstan.}
\affiliation{Fesenkov Astrophysical Institute, Observatory 23, 050020 Almaty, Kazakhstan.}

\author{Kuantay~\surname{Boshkayev}}
\email[]{kuantay@mail.ru}
\affiliation{National Nanotechnology Laboratory of Open Type,  Almaty 050040, Kazakhstan.}
\affiliation{Al-Farabi Kazakh National University, Al-Farabi av. 71, 050040 Almaty, Kazakhstan.}

\author{Talgar~\surname{Konysbayev}}
\email[]{talgar_777@mail.ru}
\affiliation{National Nanotechnology Laboratory of Open Type,  Almaty 050040, Kazakhstan.}
\affiliation{Al-Farabi Kazakh National University, Al-Farabi av. 71, 050040 Almaty, Kazakhstan.}

\author{Yergali~\surname{Kurmanov}}
\email[]{kurmanov.yergali@kaznu.kz}
\affiliation{National Nanotechnology Laboratory of Open Type,  Almaty 050040, Kazakhstan.}
\affiliation{Al-Farabi Kazakh National University, Al-Farabi av. 71, 050040 Almaty, Kazakhstan.}

\author{Guldana~\surname{Rabigulova}}
\email[]{guldanaberikhanovna@gmail.com}
\affiliation{National Nanotechnology Laboratory of Open Type,  Almaty 050040, Kazakhstan.}
\affiliation{Al-Farabi Kazakh National University, Al-Farabi av. 71,  Almaty 050040, Kazakhstan.}

\date{\today}

%\begin{abstract}

%/end{abstract}


\begin{abstract}

We explore the viability of scalar field dark matter halo models through the rotation curve analysis of the Andromeda galaxy (M31), taking into account a realistic description of its baryonic structure. The mass model includes a stellar disk described by the Freeman profile and two alternative bulge configurations: a classical single de Vaucouleurs bulge and a two-component structure consisting of inner and main bulges modeled by exponential sphere profiles. The dark matter halo is modeled using three scalar field motivated models: fuzzy dark matter (FDM), Bose–Einstein condensate and multistate scalar-field dark matter. The model parameters are determined through the Levenberg-Marquardt nonlinear least–squares fitting, and the relative performance of the models is evaluated using the Bayesian Information Criterion which allows a direct comparison with previous phenomenological halo studies performed for the same galaxy. We find that the two-bulge baryonic configuration ensures a better statistical description of the M31 rotation curve, independently of the adopted halo model. The results also suggest that, within scalar field dark matter scenarios, smooth cored halos, such as FDM, provide the most consistent description of the M31 kinematics.

\end{abstract}

\maketitle

 \section{Introduction}
 \label{sec:Intro}

The existence of dark matter (DM) is supported by a wide range of astrophysical and cosmological observations, including galaxy rotation curves (RCs), gravitational lensing, galaxy cluster dynamics, and cosmic microwave background measurements. It has been established that the observed rotational velocities of stars and gas remain approximately constant at large galactocentric distances, in clear contradiction with the Keplerian decline expected from the distribution of visible matter alone \cite{1970ApJ...159..379R,1980ApJ...238..471R,1981AJ.....86.1791B,1978AJ.....83.1026R,1979ARA&A..17..135F}. These studies indicate the presence of an extended and dominant non-luminous matter component forming galactic halos.

However, the nature of DM remains a subject of intense debate, as no suitable candidates for DM particles have been experimentally detected so far. Specifically, the existence of DM appears obvious in dynamical measurements, but not in its interactions with ordinary matter or radiation, given that ground-based laboratories have not yet been able to detect DM particles \cite{2021PhyEd..56c5011W}.

RCs represent one of the most direct probes of the mass distribution in galaxies and provide important constraints on the spatial distribution of DM \cite{2009PASJ...61..153S,2019A&ARv..27....2S,2020A&A...643A.161D}. In particular, detailed RC measurements allow one to test competing DM models by comparing their predicted velocity profiles with observations over a wide radial range extending from baryon-dominated central regions to halo-dominated outer regions.

Within the standard cosmological paradigm, DM is usually described as a cold and collisionless particle component (CDM). Numerical simulations within this framework predict cuspy density profiles such as the Navarro–Frenk–White (NFW) profile \cite{1996ApJ...462..563N}. However, observational studies of galaxy RCs often indicate the presence of central cores rather than cusps, giving rise to the well-known core–cusp problem \cite{2010AdAst2010E...5D,2014Natur.506..171P}. This discrepancy has motivated the investigation of alternative DM scenarios capable of naturally producing cored density distributions.

Among the proposed alternatives, Scalar Field Dark Matter (SFDM) models have attracted considerable attention because they provide a physically motivated mechanism for addressing several small-scale challenges of the standard CDM paradigm. In these models, DM consists of ultralight bosonic particles whose extremely small masses lead to macroscopic quantum effects on galactic scales \cite{2000PhRvL..85.1158H,2017PhRvD..95d3541H}.  One of the most important consequences of this framework is the natural formation of central density cores, which arise from quantum pressure or self-interactions of the scalar field (SF) and therefore avoid the cuspy inner profiles typically predicted by CDM simulations. In addition, they can reproduce the large-scale successes of CDM while modifying the small-scale structure of halos. In particular, SFDM halos are expected to form hierarchically in a manner similar to CDM, while condensation processes inside structures naturally lead to galactic-scale cores \cite{Robles_2015,2017MNRAS.468.3135B,2023PhRvD.108d3012F,2023MNRAS.524.4256M}

One of the most studied realizations is fuzzy dark matter (FDM), in which the wave nature of the SF leads to the formation of a central solitonic core surrounded by an extended halo \cite{2014NatPh..10..496S,2014PhRvL.113z1302S}. Other SF realizations include Bose–Einstein condensate (BEC) DM \cite{2007JCAP...06..025B} and multistate SF configurations (mSFDM) \cite{2013ApJ...763...19R}, which may produce more complex density structures depending on the superposition of quantum states. 

Testing such models against observational data is essential for assessing their viability. RC analyses serve as a particularly useful laboratory for this purpose, since SF halos often predict characteristic deviations from standard phenomenological profiles, especially in the inner regions and in the transition between baryonic and DM dominated regimes.

The Andromeda galaxy (M31) represents one of the best astrophysical systems for such studies. As the nearest massive spiral galaxy, M31 has been extensively observed across multiple wavelengths, leading to high-quality RC measurements covering both the inner galactic regions and the extended halo \cite{2015PASJ...67...75S,2010A&A...511A..89C,2006ApJ...641L.109C}. Its relatively well-constrained baryonic structure makes it an ideal candidate for our purpose.

Previous analyses of M31 rotation curves have typically employed phenomenological DM profiles such as NFW, Burkert, Einasto, and exponential sphere (ES) models \cite{2024IJMPD..3350016B}. These studies demonstrated that the inferred halo properties depend on the adopted baryonic structure. In particular, it was shown that decomposing the bulge into inner and main components can significantly improve the quality of the fits compared to a single-bulge description.

SFDM models have also been applied to the RC analysis of various galaxy samples, including low surface brightness, spiral and dwarf galaxies \cite{2023A&A...676A..63B, 2025RAA....25d5006A,2024PhRvD.110f3502A,2022PhRvD.105h3015B} (FDM), \cite{2011JCAP...05..022H,2014PhRvD..89f3507G,2015mgm..conf.1279D,2019RoAJ...29..109C,2020EPJC...80..735C,2020IJMPD..2950063C,Escamilla-Rivera_2020,2022EPJC...82..401H} (BEC), \cite{2012JCAP...01..020H,MartinezMedina2014DwarfGI,Bernal_2015,2018MNRAS.475.1447B,2021arXiv210305190S,2018IJMPD..2750031H,YeB2024FINITETE,2025JCAP...01..155B} (mSFDM). In works \cite{2020Galax...8...74B,boshkayev2020,2021MNRAS.508.1543B,boshkayev2023analysis}, phenomenological DM models were analyzed using Levenberg-Marquardt nonlinear fitting procedure within the RCs of different galaxies.

While SFDM models have been successfully applied to the RC analysis of various galaxies, the level of detail in the adopted baryonic descriptions varies significantly between studies. This makes it useful to further examine how SF motivated halos perform when combined with detailed baryonic mass models, and to assess their statistical behavior in comparison with commonly used phenomenological profiles.

The aim of the present work is to address this question by testing several SFDM halo models against the observed RC of M31. Firstly, we assess whether SFDM models can reproduce the observed M31 RC when baryonic components are taken into account. Secondly, we examine whether the statistical preference for particular halo models reflects intrinsic properties of the DM sector or rather the structural complexity of the baryonic components.

In particular, we investigate three SF halo realizations: the FDM model, the BEC profile, and the mSFDM model. The baryonic mass distribution is modeled using two alternative descriptions: a classical single-bulge model based on the de Vaucouleurs profile and a two-component bulge decomposition motivated by observational studies of galactic inner structure. The model parameters are determined through the Levenberg-Marquardt nonlinear fitting procedures, and the relative performance of the models is evaluated using the Bayesian Information Criterion.

This paper is organized as follows. In Section \ref{sec:Structure} we describe the adopted mass model and the density profiles used for the baryonic and DM components. Section \ref{sec:Data_Methods} presents the observational data and the fitting methodology. The fitting results are summarized in \ref{sec:Results}. In Section \ref{sec:Discussion} we discuss the physical implications of the obtained results and compare them with previous analyses. Finally, Section \ref{sec:Conclusion} summarizes our main conclusions and perspectives.

\section{Galaxy structure}\label{sec:Structure}

In order to interpret the observed circular velocities, it is necessary to construct a dynamical model that accounts for both the luminous baryonic matter and the DM halo. The Andromeda galaxy (M31) is among the best studied spiral galaxies, for which the baryonic structure has been investigated in considerable detail. The system exhibit a complex inner morphology characterized by a central bulge and an extended stellar disk, embedded within a massive DM halo that dominates the gravitational potential at large radii. A realistic modeling of the baryonic distribution is therefore an essential ingredient in the interpretation of the RC data, since the inner regions of galaxies are primarily influenced by the stellar components.

In this work we construct a mass model that incorporates the main baryonic constituents of the galaxies together with a DM halo component. In the following subsections we describe in detail the density profiles adopted for the bulge, disk, and DM halo, and derive the corresponding contributions to the RCs.

\subsection{The bulge}\label{sec:bulge}

The stellar bulge is described using two ways in order to account for possible structural differences in the inner regions of the galaxies. In the first approach the bulge is modeled by the classical de Vaucouleurs profile \cite{1958ApJ...128..465D}, which has been widely employed to represent spheroidal stellar distributions in massive galaxies. The surface brightness distribution is given by
\begin{equation}\label{Vaucouleurs_prof}
    \Sigma_b(x) = \Sigma_{bc}\text{exp}\left[\kappa\left(1-x^{1/4}\right)\right],
\end{equation}
where $\kappa=7.6695$, $x=r/r_b$, $\Sigma_{bc}$ is the central value of the surface mass density, $r_b$ is the scale radius of the bulge. The total bulge mass is defined as
\begin{equation}
M_{bt}=2\pi\int_0^{\infty}\Sigma_b(r)r~dr = \eta r_b^2\Sigma_{bc},
\end{equation}
where $\eta=22.665$ is a dimensionless constant. The mass enclosed within a sphere of radius $r$ is given by
\begin{equation}
    M_{bt}(r) = 4\pi\int_0^r\rho_b(y)y^2~dy,
\end{equation}
where the density is calculated as
\begin{equation}
    \rho_b(y)=\frac{1}{\pi}\int_y^{\infty}\frac{d\Sigma_b(x)}{dx}\frac{dx}{\sqrt{x^2-y^2}}.
\end{equation}

In addition, we adopt a double bulge model, motivated by observational studies indicating that de Vaucouleurs law fails to fit the bulge RC of the Milky Way galaxy. In this case the bulge can be devided into inner and main bulges, both modeled by Exponential Sphere (ES) profiles \cite{2013PASJ...65..118S}:
\begin{equation}\label{ES_profile}
    \rho(x)=\rho_0 \text{exp}(-x),
\end{equation}
where $x=r/r_0$. The corresponding mass contained in a sphere with a radius $r$ reads
\begin{equation}
    M(r)=M_0\left[1-\text{exp}(-x)\left(1+x+x^2/2\right)\right],
\end{equation}
and the total mass $M_0$ is
\begin{equation}
    M_0=4\pi\int_0^{\infty}r^2\rho(r)~dr=8\pi r_0^3\rho_0.
\end{equation}
One should emphasize that Eq.~\eqref{Vaucouleurs_prof} represents the surface density, while Eq.~\eqref{ES_profile} includes the volume density, and therefore we
can't compare these two profiles to each other.

\subsection{The galactic disk}\label{sec:disk}

The galactic disk constitutes the dominant baryonic component at intermediate galactocentric radii in spiral galaxies. The radial distribution of stellar matter in galactic disks is well described by an exponential profile originally introduced by Freeman \cite{1970ApJ...160..811F}. Accordingly, the surface mass density of the disk is written as
\begin{equation}\label{Freeman_profile}
    \Sigma_d(x)=\Sigma_{dc}\text{exp}(-x),
\end{equation}
where $x=r/r_d$, $\Sigma_{dc}$ is the central value of the surface density and $r_d$ is the disk scale radius. The mass enclosed within a radius $r$ is
\begin{equation}
    M_d(r)=M_{dt}\left[1-\text{exp}(-x)(1+x)\right],
\end{equation}
where the total mass of a disk is given by
\begin{equation}
    M_{dt}=2\pi\int_0^{\infty}\Sigma_d(r)r~dr = 2\pi r_d^2 \Sigma_{dc}.
\end{equation}
The rotation curve velocity for a thin exponential disk, according to \cite{1987gady.book.....B}, can be expressed as
\begin{equation}\label{vel_disk}
    v_d(y)=\sqrt{\frac{2GM_{dt}}{r_d}y^2\left[I_0(y)K_0(y)-I_1(y)K_1(y)\right]},
\end{equation}
where $y=r/(2r_d)$ and $I_i$, $K_i$ ($i=1, 2$) are the modified Bessel functions of the first and second kind. respectively.

\subsection{Fuzzy DM model}\label{sec:FDM}

Apart from the baryonic component, the mass model includes a DM halo responsible for maintaining the approximately flat behavior of the RCs at large distances.

The FDM model is described at the fundamental level by the coupled Schrödinger–Poisson system, which governs the evolution of a self-gravitating ultralight scalar field.

The density profile of the solitonic core is commonly approximated by \cite{2014NatPh..10..496S,2014PhRvL.113z1302S}
\begin{equation}\label{FDM_profile}
    \rho_{FDM}(r) = \frac{\rho_c}{\left[1+0.091\left(\frac{r}{r_c}\right)^2\right]^8},
\end{equation}
where $\rho_0$ the central density and $r_0$ is the core radius. The core radius is inversely related to the particle mass which is of order 
$m\sim10^{-22}$ eV. At galactic scales the associated de Broglie wavelength becomes significant, leading to a wave-like behavior of the DM distribution. Numerical simulations of this model predict the formation of a central solitonic core surrounded by an extended halo.

\subsection{BEC SFDM model}\label{sec:BECSFDM}
Another important scenario arises when ultralight bosonic particles form a Bose–Einstein condensate (BEC) on galactic scales. In this framework the DM halo can be described as a self-gravitating condensate governed by the Gross–Pitaevskii equation coupled to the Poisson equation.

Under the Thomas–Fermi approximation, the equilibrium density distribution of the condensate takes the form \cite{2007JCAP...06..025B}:
\begin{equation}\label{BEC_profile}
    \rho_{BEC}(r)=\frac{\rho_0 \sin(kr)}{kr},
\end{equation}
where $\rho_c$ is the central density and $k=\pi/r_0$  is related to the halo radius $r_0$. This profile was derived in the context of model in which the bosonic particles exhibit repulsive self-interactions. The BEC profile predicts a central core and therefore naturally avoids the cusp predicted by conventional CDM halos.

\subsection{Multistate SFDM model}\label{sec:mSFDM}

The multistate configuration can be interpreted as a finite-temperature extension, in which the halo is not fully condensed into the ground state but instead contains a superposition of excited states. The number of contributing states determines the level of structural complexity of the halo.

Following the formulation proposed by Robles and Matos, the density distribution can be written as a superposition of several states \cite{2013ApJ...763...19R}
\begin{equation}\label{mSFDM_profile}
    \rho_{mSFDM}(r)=\sum_j\rho_0^j\left[\frac{\sin \left(k_jr\right)}{k_jr}\right]^2,
\end{equation}
where $j=1,2,3,...$ is the number of exited states required to fit the distribution, $\rho_0^j$ is a central density of a single $j$-th state, and $k_jr_0 = j\pi$. As a result, this model is capable of reproducing more intricate radial features, although the presence of oscillatory patterns in the density profile may affect its statistical performance when compared to smoother halo models.

\section{Data and methods}\label{sec:Data_Methods}
\subsection{Rotation Curve data}

The observational RC data of the Andromeda galaxy (M31) used in this work were taken from the compilation presented by Sofue \cite{2015PASJ...67...75S}, which represents one of the most detailed and widely used datasets for this galaxy. The dataset combines optical and 21-cm observations and covers the galactocentric region from the inner bulge up to the outer halo, thus providing a reliable basis for testing mass distribution models across all galactic components.

The RC consists of $N=46$ velocity measurements with corresponding radial distances and observational uncertainties, which are explicitly taken into account in the fitting procedure. The inclusion of observational errors allows a statistically consistent determination of the model parameters and ensures an objective comparison between different halo realizations.

Following the methodology adopted in previous studies of M31 \cite{2024IJMPD..3350016B}, we do not perform separate fits of individual galactic components. Instead, we simultaneously fit all baryonic and DM contributions using the full set of RC data points.

The adopted RC dataset spans the full structural extent of the galaxy, including the bulge-dominated central region, the disk-dominated intermediate region, and the DM dominated halo region extending up to several hundred kiloparsecs. Such radial coverage is essential for distinguishing between competing DM models since different halo profiles typically produce similar behaviour in the inner regions but diverge significantly at large radii.

\subsection{Methods}
The observed RCs of spiral galaxies ensure a direct measure of the gravitational potential generated by the total mass distribution. Under the assumption of dynamical equilibrium and approximate circular motion, the circular velocity at radius $r$ is determined by the derivative of the total gravitational potential $\Phi(r)$. In the framework of a multi-component mass model, the total circular velocity is expressed as the quadratic sum of the contributions from the baryonic and DM components:
\begin{equation}\label{total_rot_vel}
    v^2(r)=v_b^2(r)+v_d^2(r)+v_h^2(r),
\end{equation}
where $v_b(r)$, $v_d(r)$ and $v_h(r)$ are the circular velocities associated with the bulge, stellar disk and DM halo, respectively.

Since the detailed structure of the central regions of both galaxies remains subject to observational uncertainties\footnote{}, we consider two alternative descriptions of the bulge component in order to evaluate the sensitivity of the DM parameters to the adopted baryonic model. In the first approach, the bulge is treated as a single stellar component described by the de Vaucouleurs profile. In the second approach, motivated by observational evidence for multiple stellar structures in galactic centers, the bulge is decomposed into two components: an inner bulge and a main bulge, both described by exponential density profiles but characterized by different scale radii and central densities. In the last case, the total circular velocity is given by
\begin{equation}
    v^2(r)=v_{ib}^2(r)+v_{mb}^2(r)+v_d^2(r)+v_h^2(r),
\end{equation}
where the velocities $v_{ib}$ and $v_{mb}$ refer to the inner bulge and main bulge, correspondingly.

For each galaxy, the circular velocity profile is constructed by computing the individual contributions from the adopted density profiles described in Sec.~\ref{sec:Structure}. The bulge and halo contributions are calculated from the enclosed mass profiles via
\begin{equation}
    v(r)=\sqrt{\frac{GM(r)}{r}} = \sqrt{r\frac{\partial\Phi(r)}{\partial r}},
\end{equation}
while the disk contribution is calculated using the analytical expression for the exponential surface density profile (Eq.~\eqref{vel_disk}). The mass distribution in a halo is defined by integrating the density profile
\begin{equation}
    M(r)=\int_0^r4\pi x^2\rho(x)dx.
\end{equation}

%The total RC is then obtained by summing all contributions at each radial point corresponding to the observational data. This approach ensures that the gravitational influence of all components is treated self-consistently within a unified dynamical model.
The free parameters of the model will be obtained by fitting the theoretical RC to the observational data. In contrast to approaches where the baryonic and DM components are fitted separately, in this work we perform an unified fit in which all model parameters are determined simultaneously. This method reduces degeneracies that may arise when components are treated independently.

The best-fit parameters are determined using the Levenberg-Marquadt nonlinear least-squares procedure \cite{levenberg1944method,marquardt1963algorithm} based on the minimization of the $\chi^2$ function:
\begin{equation}
    \chi^2 = \sum_{i=1}^N\left[\frac{v_i^{obs}-v(r)}{\sigma_{v,i}^{obs}}\right]^2,
\end{equation}
where $N$ is the number of data points, $v_i^{obs}$ is the observed circular velocity, $\sigma_{v,i}^{obs}$ is the observational uncertainty, and $v(r)$ is the model prediction. Initial parameter values are selected based on typical ranges reported in \cite{2015PASJ...67...75S}.

Since the considered halo models contain different numbers of free parameters and different functional forms, we compare their performance using the Bayesian Information Criterion (BIC). The BIC is defined as:
\begin{equation}\label{BIC}
    BIC=\chi^2+k \ln N,
\end{equation}
where $k$ is the number of free parameters in a model. The BIC is useful for models with a larger number of parameters, and the preferred model is that one with the lowest value of BIC ($BIC_0$). When comparing models, the difference
\begin{equation}\label{difBIC}
    \Delta BIC = BIC-BIC_0
\end{equation}
can be interpreted as:
\begin{itemize}
\item{$\Delta BIC \in [0,2]$} - weak evidence;
\item{$\Delta BIC \in (2,6]$} - positive evidence;
\item{$\Delta BIC>6$} - strong evidence.
\end{itemize}

\section{Theoretical results}\label{sec:Results}

In this section we present the results of fitting the observed rotation curves of the M31 galaxy within the framework of specified scenarios. The resulting best–fit RCs and the corresponding parameters are summarized in the Figures \ref{fig:RC_M31_FDM}-\ref{fig:RC_M31_mSFDM} and Tables \ref{tab:fit_M31_1bulge}-\ref{tab:fit_M31_2bulges} below.

\begin{figure*}[ht]
\begin{minipage}{0.5\linewidth}
\center{\includegraphics[width=1\linewidth]{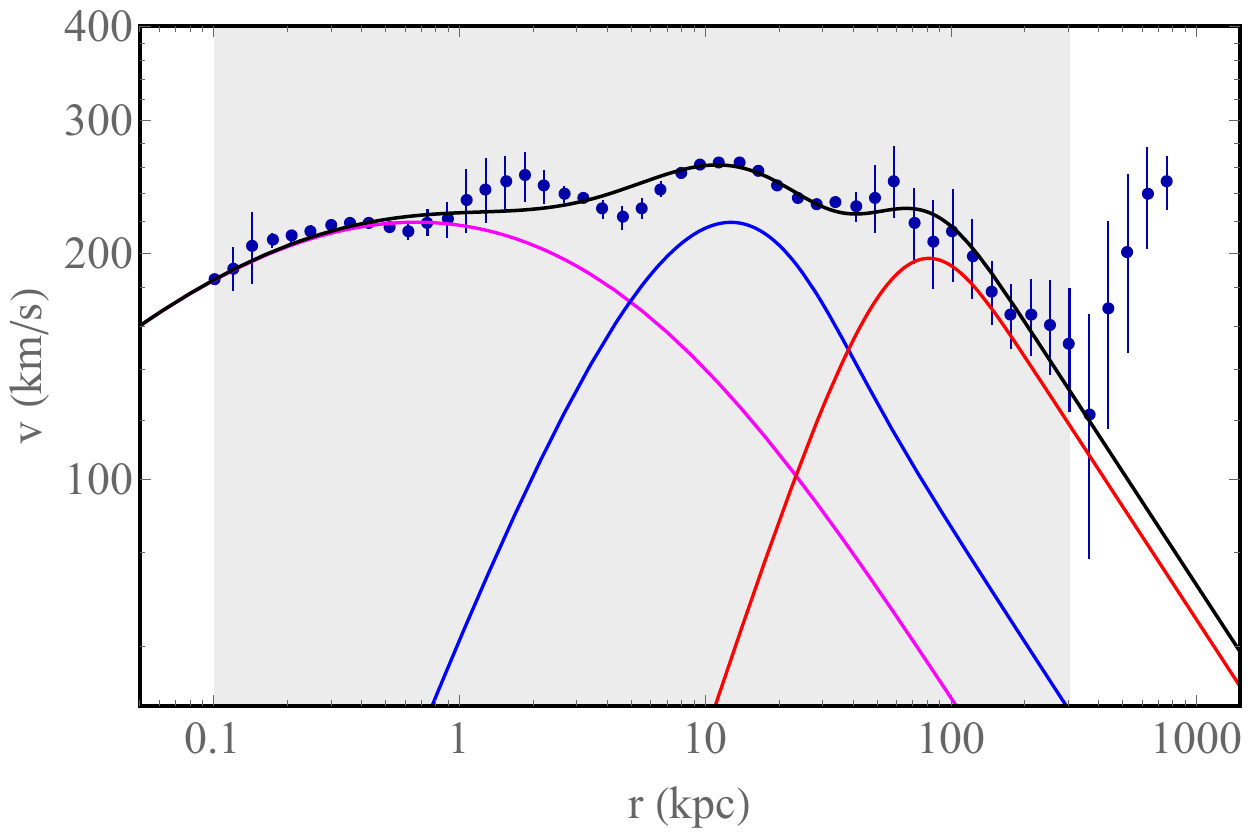}\\ }
\end{minipage}
\hfill 
\begin{minipage}{0.49\linewidth}
\center{\includegraphics[width=1.\linewidth]{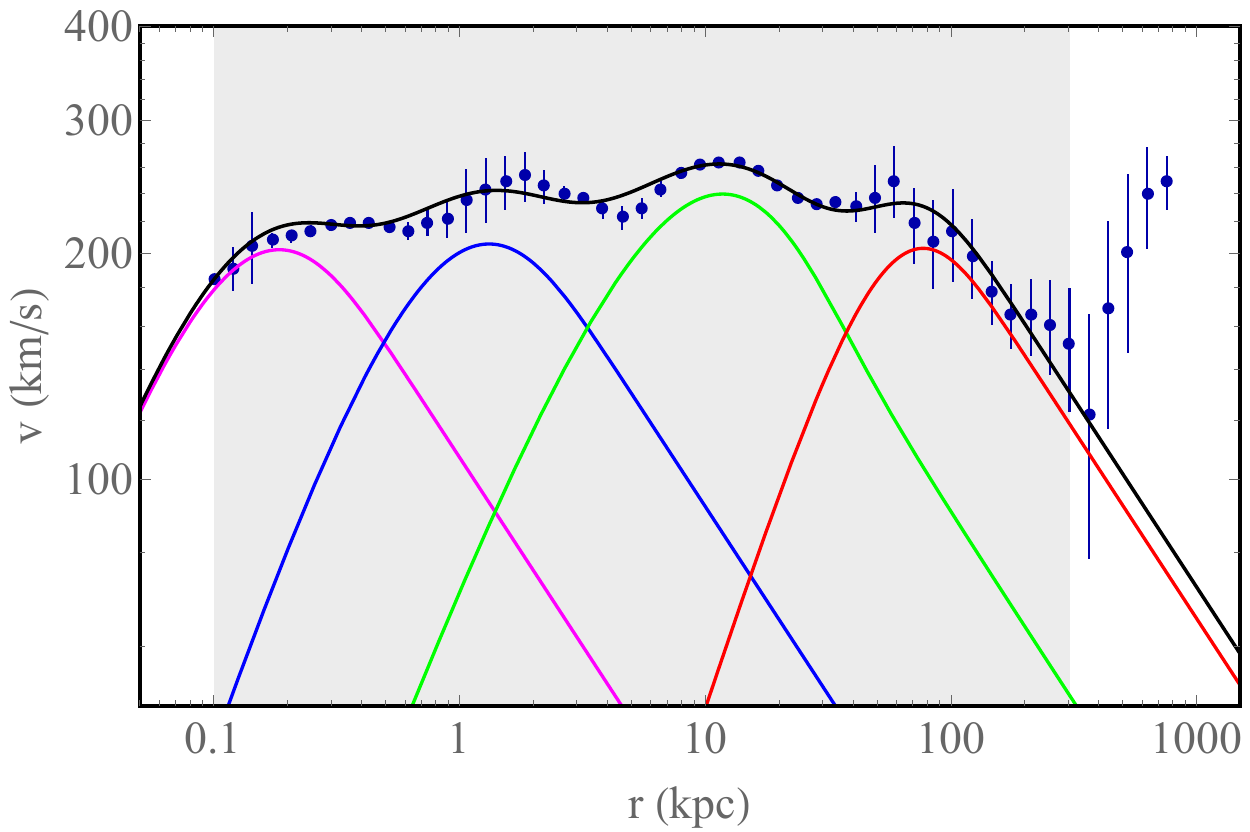}\\ }
\end{minipage}
\caption{Reconstructed RC of M31. The shaded area marks the data (black dots with error bars) used in the fitting. Left panel: Best-fit velocity profile (solid black) composed of single bulge (de Vauculeurs, purple), disk (Freeman, blue) and halo (FDM, red) curves. Right panel: Best-fit velocity profile (solid black) composed of inner bulge (ES, purple), main bulge (blue), disk (Freeman, green) and halo (FDM, red) curves.}
\label{fig:RC_M31_FDM}
\end{figure*}

Figure \ref{fig:RC_M31_FDM} shows the fits obtained with the FDM halo for the two adopted baryonic mass models. In the left panel, where the bulge is described by a single de Vaucouleurs component, the inner rise of the total circular velocity is mainly controlled by the bulge, while the disk contribution dominates the intermediate radial range. The FDM halo contribution becomes significant in the outer region and provides the additional support required to maintain the observed velocities at large radii. In the right panel, where the bulge is decomposed into inner and main ES components, the same global shape of the observed RC is reproduced, but the central baryonic contribution is redistributed more realistically between a compact inner structure and a more extended main bulge. This decomposition leads to a smoother transition between the central and intermediate regions and yields the overall best statistical fit among all models considered for M31, with the lowest BIC value reported in Table \ref{tab:fit_M31_2bulges}.

\begin{figure*}[ht]
\begin{minipage}{0.49\linewidth}
\center{\includegraphics[width=1\linewidth]{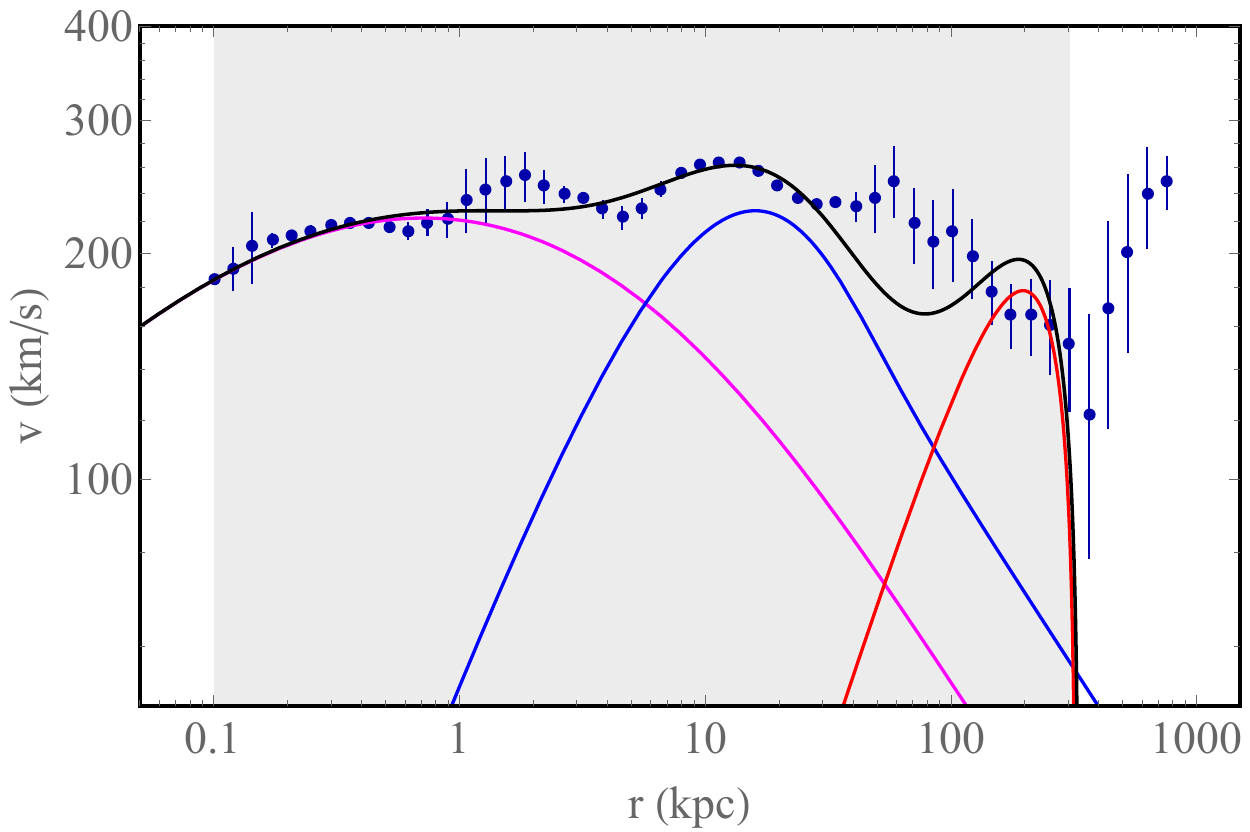}\\ }
\end{minipage}
\hfill 
\begin{minipage}{0.495\linewidth}
\center{\includegraphics[width=1.\linewidth]{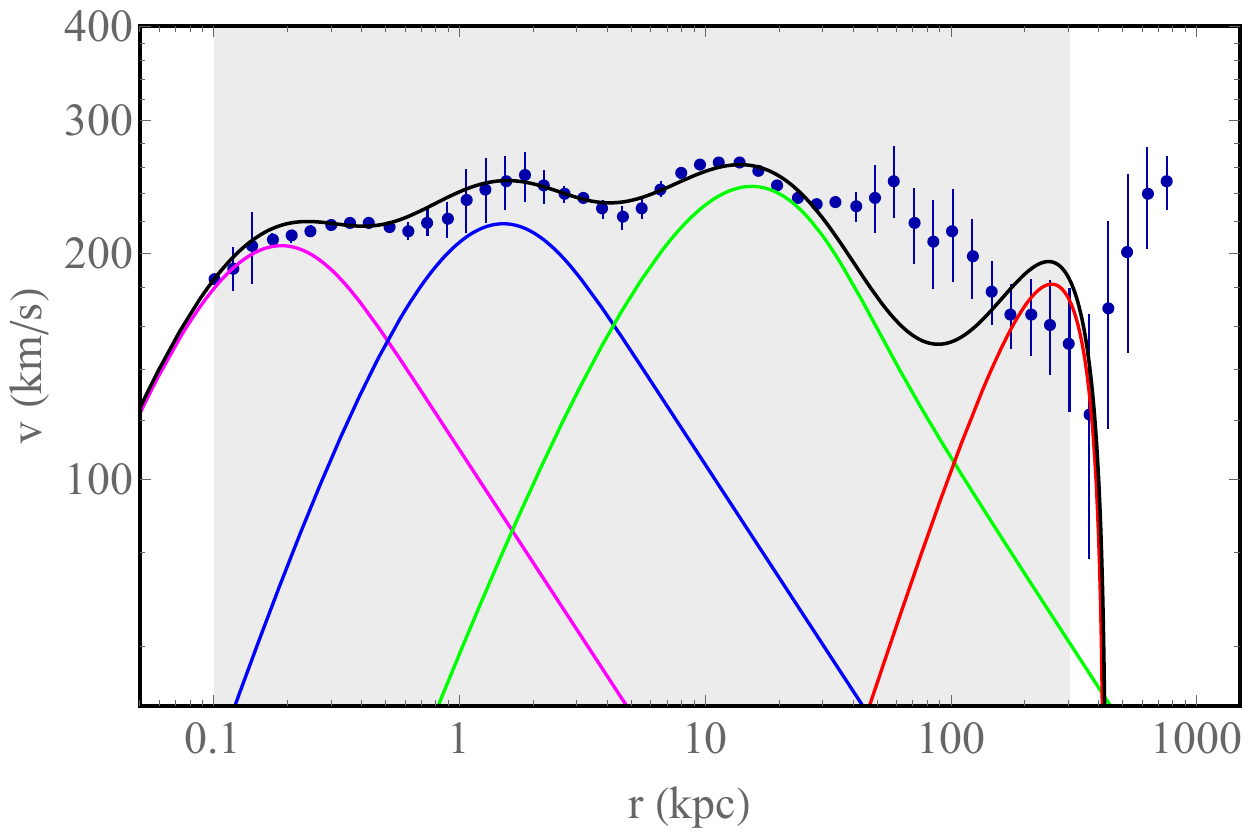}\\ }
\end{minipage}
\caption{Reconstructed RC of M31. The shaded area marks the data (black dots with error bars) used in the fitting. Left panel: Best-fit velocity profile (solid black) composed of single bulge (de Vauculeurs, purple), disk (Freeman, blue) and halo (BEC, red) curves. Right panel: Best-fit velocity profile (solid black) composed of inner bulge (ES, purple), main bulge (blue), disk (Freeman, green) and halo (BEC, red) curves.}
\label{fig:RC_M31_Harko}
\end{figure*}

Figure \ref{fig:RC_M31_Harko} presents the corresponding results for the BEC halo. In both baryonic descriptions, the model can reproduce the general trend of the observed RC in the inner and intermediate parts of the galaxy. However, compared with the FDM case, the BEC halo requires a more extended characteristic radius and lower central density, as seen from Tables \ref{tab:fit_M31_1bulge} and \ref{tab:fit_M31_2bulges}. Its statistical performance is clearly worse, especially in the two-bulge case, where the BIC becomes significantly larger than that of the FDM model. This suggests that the BEC profile is less efficient in matching the detailed radial structure of the M31 RC.

\begin{figure*}[ht]
\begin{minipage}{0.49\linewidth}
\center{\includegraphics[width=1\linewidth]{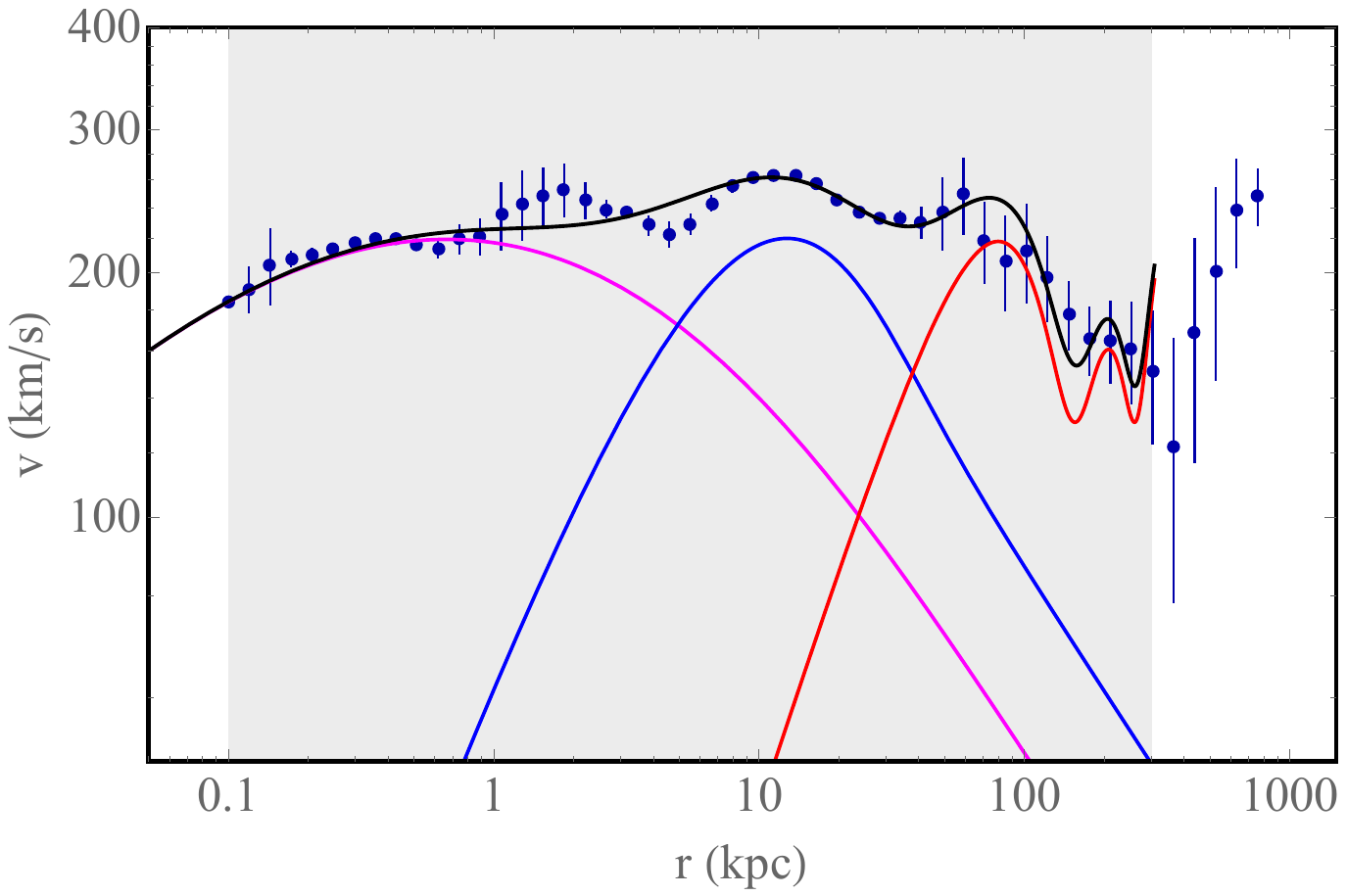}\\ }
\end{minipage}
\hfill 
\begin{minipage}{0.495\linewidth}
\center{\includegraphics[width=1.\linewidth]{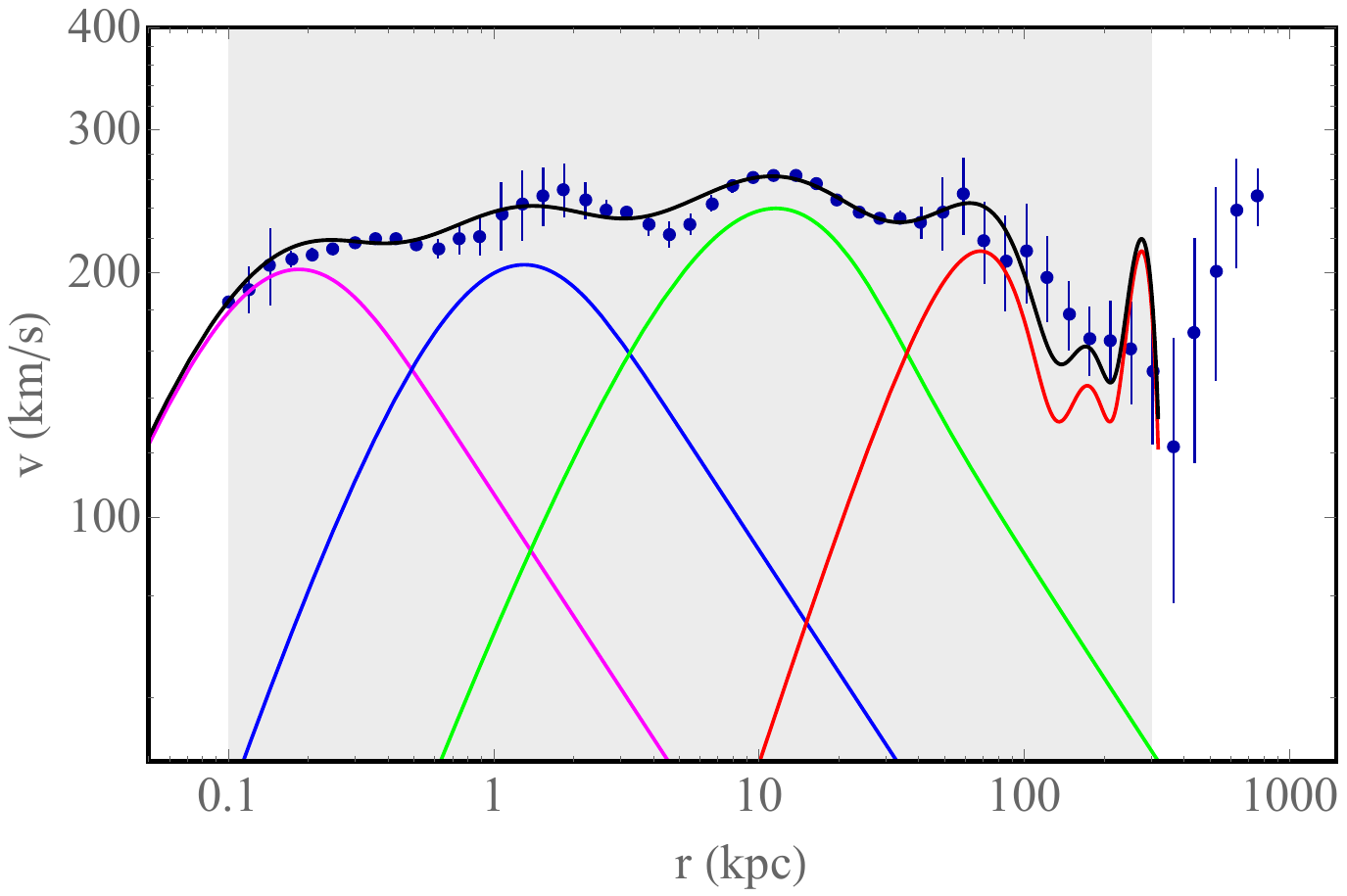}\\ }
\end{minipage}
\caption{Reconstructed RC of M31. The shaded area marks the data (black dots with error bars) used in the fitting. Left panel: Best-fit velocity profile (solid black) composed of single bulge (de Vauculeurs, purple), disk (Freeman, blue) and halo (BEC, red) curves. Right panel: Best-fit velocity profile (solid black) composed of inner bulge (ES, purple), main bulge (blue), disk (Freeman, green) and halo (mSFDM, red) curves.}
\label{fig:RC_M31_mSFDM}
\end{figure*}

Figure \ref{fig:RC_M31_mSFDM} shows the fit obtained with the mSFDM halo. In comparison to the FDM case, the mSFDM contribution reflects a more structured halo configuration associated with the presence of multiple states. Nevertheless, despite this additional freedom, the fit does not outperform the FDM model. Its BIC values remain higher than those of the FDM fits for both baryonic descriptions, although they are lower than the BEC result in the two-bulge case. This indicates that the multistate configuration is capable of reproducing the observed RC, but that the extra halo complexity is not strongly favored by the present M31 data once the baryonic components are fitted simultaneously.

\begin{figure*}[ht]
\begin{minipage}{0.49\linewidth}
\center{\includegraphics[width=1\linewidth]{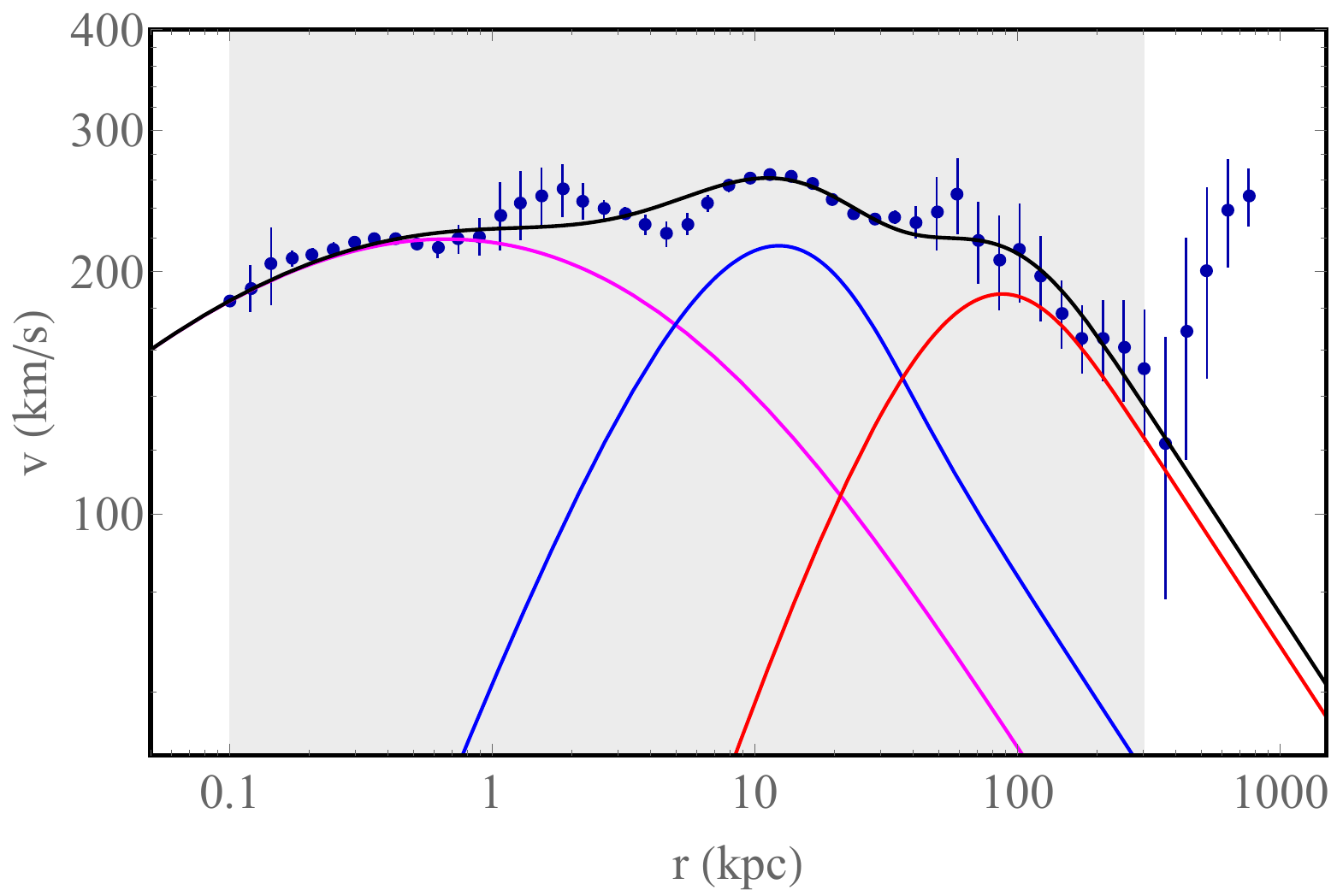}\\ }
\end{minipage}
\hfill 
\begin{minipage}{0.495\linewidth}
\center{\includegraphics[width=1.\linewidth]{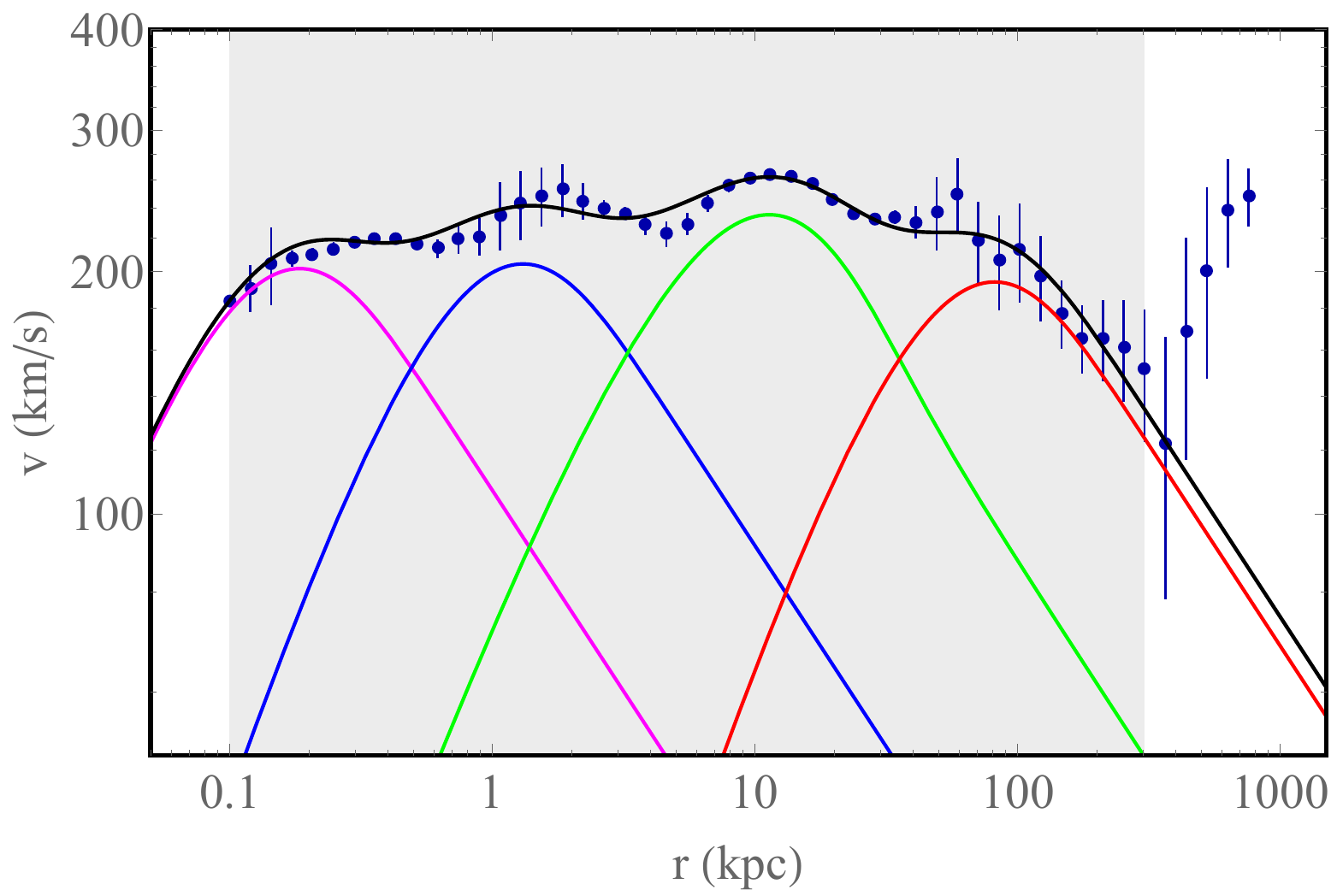}\\ }
\end{minipage}
\caption{Reconstructed RC of M31. The shaded area marks the data (black dots with error bars) used in the fitting. Left panel: Best-fit velocity profile (solid black) composed of single bulge (de Vauculeurs, purple), disk (Freeman, blue) and halo (ES, red) curves. Right panel: Best-fit velocity profile (solid black) composed of inner bulge (ES, purple), main bulge (blue), disk (Freeman, green) and halo (ES, red) curves.}
\label{fig:RC_M31_ES}
\end{figure*}

An additional reference model is provided by the exponential-sphere (ES) halo, whose best-fit rotation curves are shown in Fig.~\ref{fig:RC_M31_ES}. This profile is included as a representative phenomenological halo in order to provide a direct comparison to the bosonic and DM scenarios. In previous analyses of M31 and the Milky Way performed within the same fitting framework \cite{2024IJMPD..3350016B}, the ES profile yielded the lowest BIC among several phenomenological halos, which motivates its use here as a benchmark model.

%%%%%%%%%%%%%%%%%%%%%%%%%%

\begin{table*}[t]
\centering
\caption{Best-fit RC parameters for M31 obtained by considering a single bulge (de Vaucouleurs), disk (Freeman) and halo models (FDM, BEC, mSFDM). The halo mass has been calculated within $r_c=200$ kpc. The $\Delta \mathrm{BIC}$ value is computed with respect to the minimum BIC in Table \ref{tab:fit_M31_2bulges}.}
\label{tab:fit_M31_1bulge}
\renewcommand{\arraystretch}{1.3}
\setlength{\tabcolsep}{8pt}

\begin{tabular}{ccccccccc}

\hline\hline

\multicolumn{2}{c}{Bulge} & 
\multicolumn{2}{c}{Disk} & 
\multicolumn{3}{c}{Halo} & 
\multirow{2}{*}{BIC} &
\multirow{2}{*}{$\Delta \mathrm{BIC}$} \\

\cline{1-7}

$M_b$ & $r_b$ &
$M_d$ & $r_d$ &
$\rho_0$ & $r_0$ & $M_h$ &
& \\

$(10^{11}M_\odot)$ &
$(\mathrm{kpc})$ &
$(10^{11}M_\odot)$ &
$(\mathrm{kpc})$ &
$(10^{-3}M_\odot/\mathrm{pc}^3)$ &
$(\mathrm{kpc})$ &
$(10^{11}M_\odot)$ &
& \\

\hline

\hline

\multicolumn{2}{c}{de Vaucouleurs} &
\multicolumn{2}{c}{Freeman} &
\multicolumn{3}{c}{ES} &
& \\

 0.613 & 4.051 & 1.598 & 5.741 & 2.504 & 25.860 & 10.885 & 177.8 & 6.5 \\

 \hline

\multicolumn{2}{c}{de Vaucouleurs} &
\multicolumn{2}{c}{Freeman} &
\multicolumn{3}{c}{FDM} &
& \\

0.616 & 4.061 & 1.715 & 5.930 & 1.180 & 41.611 & 9.853 & 175.6 & 4.3 \\

\hline

\multicolumn{2}{c}{de Vaucouleurs} &
\multicolumn{2}{c}{Freeman} &
\multicolumn{3}{c}{BEC} &
& \\

 0.677 & 4.354 & 2.313 & 7.448 & 0.107 & 225.763 & 14.731 & 210.4 & 24.8 \\

\hline

\multicolumn{2}{c}{de Vaucouleurs} &
\multicolumn{2}{c}{Freeman} &
\multicolumn{3}{c}{mSFDM} &
& \\

 0.615 & 4.057 & 1.729 & 5.943 & $\rho_0^1$: 0.112 & 416.160 & 1.179 & 200.2 & 24.6 \\
 &  &  &  & $\rho_0^3$: 0.245 &  &  &  & \\

\hline\hline

\end{tabular}
\end{table*}

The best-fit parameters listed in Table \ref{tab:fit_M31_1bulge} correspond to the mass model with a single de Vaucouleurs bulge, a Freeman disk, and one of the three halo profiles. For this baryonic prescription, the FDM model yields the best fit, with $BIC=175.6$ and $\Delta BIC=4.3$, whereas the BEC and mSFDM models give significantly larger values, $210.4$ and $200.2$, respectively. The fitted FDM halo is characterized by a comparatively high central density, $\rho_0=1.180\times10^{-3}M_{\odot}/pc^3$, and a moderate scale radius, $r_0=41.611$ kpc, which together produce a halo mass of $M_h=9.853\times10^{11}M_{\odot}$ within 200 kpc.
%%%%%%%%%%%%%%%%%%%%%%%%%%%%%%

\begin{table*}[t]
\centering
\caption{Best-fit RC parameters for M31 obtained by considering inner (ES) and main (ES) bulges, disk (Freeman) and halo models (FDM, BEC, mSFDM). The halo mass has been calculated within $r_c=200$ kpc. The $\Delta \mathrm{BIC}$ value is computed with respect to the minimum BIC in this Table.}
\label{tab:fit_M31_2bulges}
\renewcommand{\arraystretch}{1.3}
\setlength{\tabcolsep}{7pt}
\begin{tabular}{ccccccccccc}
\hline\hline
\multicolumn{2}{c}{Inner Bulge} & 
\multicolumn{2}{c}{Main Bulge} & 
\multicolumn{2}{c}{Disk} & 
\multicolumn{3}{c}{Halo} & 
\multirow{2}{*}{BIC} &
\multirow{2}{*}{$\Delta \mathrm{BIC}$} \\
\cline{1-9}
$M_{ib}$ & $r_{ib}$ & $M_{mb}$ & $r_{mb}$ & $M_d$ & $r_d$ & $\rho_0$ & $r_0$ & $M_h$ & & \\
$(10^{11}\,M_\odot)$ & $(\mathrm{kpc})$ & $(10^{11}\,M_\odot)$ & $(\mathrm{kpc})$ & $(10^{11}\,M_\odot)$ & $(\mathrm{kpc})$ & $(10^{-3}\,M_\odot/\mathrm{pc}^3)$ & $(\mathrm{kpc})$ & $(10^{11}\,M_\odot)$ & & \\

\hline

\multicolumn{2}{c}{ES} & \multicolumn{2}{c}{ES} & \multicolumn{2}{c}{Freeman} & \multicolumn{3}{c}{ES} & & \\
0.0266 & 0.0546 & 0.194 & 0.387 & 1.756 & 5.280 & 3.066 & 24.191 & 10.908 & 173.9 & 2.7 \\

\hline

\multicolumn{2}{c}{ES} & \multicolumn{2}{c}{ES} & \multicolumn{2}{c}{Freeman} & \multicolumn{3}{c}{FDM} & & \\
0.0267 & 0.0547 & 0.197 & 0.391 & 1.887 & 5.488 & 1.396 & 34.467 & 9.935 & 171.3 & 0 \\

\hline

\multicolumn{2}{c}{ES} & \multicolumn{2}{c}{ES} & \multicolumn{2}{c}{Freeman} & \multicolumn{3}{c}{BEC} & & \\
0.0279 & 0.0558 & 0.255 & 0.447 & 2.598 & 7.210 & 0.0651 & 294.993 & 13.395 & 223.191 & 51.9 \\

\hline

\multicolumn{2}{c}{ES} & \multicolumn{2}{c}{ES} & \multicolumn{2}{c}{Freeman} & \multicolumn{3}{c}{mSFDM} & & \\
0.0265 & 0.0546 & 0.193 & 0.386 & 1.862 & 5.398 & $\rho_0^1$: 0.176 & 345.936 & 8.340 & 195.9 & 24.6 \\
 &  &  &  &  &  & $\rho_0^3$: 0.289 &  &  &  &  \\

\hline\hline
\end{tabular}
\end{table*}

Table \ref{tab:fit_M31_2bulges} summarizes the results obtained when the bulge is decomposed into inner and main ES components. This more detailed baryonic description leads to a clear improvement for the FDM halo, which now provides the global best fit for M31 with $BIC=171.3$. The fitted inner bulge is very compact, with $M_{ib}=0.0267\times 10^{11}M_{\odot}$ and $r_{ib}=0.0547$ kpc, while the main bulge is more extended, with $M_{mb}=0.197\times 10^{11}M_{\odot}$ and $r_{mb}=0.391$ kpc. The FDM halo is described by $\rho_0=1.396\times 10^{-3}M_{\odot}/pc^3$, $r_0=34.467$ kpc and $M_h=9.935\times 10^{11}M_{\odot}$. Relative to the single-bulge fit, the preferred FDM halo becomes slightly denser and more compact.

The two-bulge results also show that the BEC and mSFDM fits remain less favorable than the FDM one. In particular, the BEC model yields the largest BIC value, 223.191, corresponding to $\Delta BIC=51.9$, which constitutes strong evidence against this profile relative to the best-fit model within the considered set. Its fitted parameters again point to a very extended and diffuse halo. The mSFDM model performs better than BEC, with $BIC=195.9$ and $\Delta BIC=24.6$, but it is still disfavored compared to FDM. Therefore, the statistical comparison demonstrates that the FDM halo consistently provides the most successful description, particularly when the baryonic component is modeled by separate inner and main bulges.

As also can be seen, the ES halo reproduces the overall shape of the observed RC with a quality comparable to the best bosonic model. In the single-bulge configuration, the ES fit yields $BIC = 177.8$, which is slightly worse than the FDM result ($BIC = 175.6$) but significantly better than the BEC and mSFDM models. A similar trend is observed in the two-bulge configuration, where the ES halo remains competitive ($BIC = 173.9$) but is still slightly disfavored compared to the FDM model ($BIC = 171.3$). These results indicate that the FDM halo achieves a level of agreement with the M31 RC comparable to, and slightly better than, the best phenomenological halo considered here.

Then, using the obtained fitting parameters, we can compute the logarithmic slope $\gamma(r)$ which is defined as
\begin{equation}
    \gamma(r)=\frac{d\ln \rho(r)}{d\ln r} = \frac{r}{\rho(r)}\frac{d\rho(r)}{dr},
\end{equation}
and quantifies the radial variation of the applied DM density profiles.
\begin{figure*}[ht]
\begin{minipage}{0.49\linewidth}
\center{\includegraphics[width=1\linewidth]{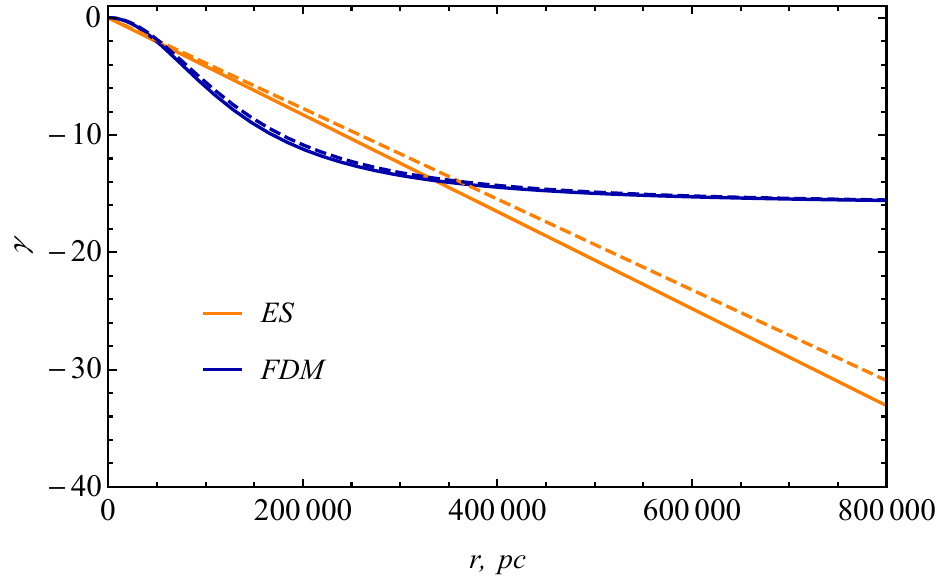}\\ }
\end{minipage}
\hfill 
\begin{minipage}{0.495\linewidth}
\center{\includegraphics[width=1.\linewidth]{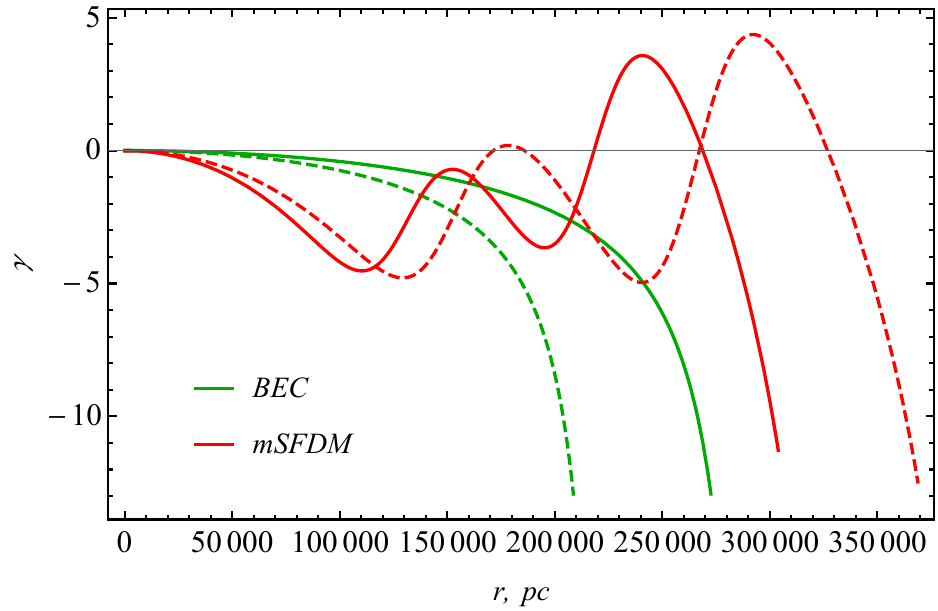}\\ }
\end{minipage}
\caption{Logarithmic slope for the considered halo models in M31. Left: ES (orange), FDM (blue). Right: BEC (green) and mSFDM (red). Solid lines correspond to mass model fitted with inner and main bulges; dashed lines correspond to the case with a single bulge.}
\label{fig:slopes_halos}
\end{figure*}

Figure \ref{fig:slopes_halos} shows the radial behaviour of the logarithmic density slopes for the FDM, BEC and mSFDM halo models obtained from the RC fitting of M31. Solid curves correspond to the mass model including inner and main bulges, while dashed curves represent the case where the bulge is treated as a single component. The left panel shows the behaviour of the ES and FDM halo slopes, both of which are smooth and monotonic. The ES profile demonstrates a linear steepening of the density slope with increasing radius, which reflects the exponential decrease of the density distribution. In contrast, the FDM model shows a much slower variation of the slope, remaining nearly constant at large radii. This behaviour is characteristic of cored SF halos, where the density decreases more gradually than in exponential profiles. The small difference between the solid and dashed curves indicates that the inferred slope of these halos is only weakly sensitive to the adopted baryonic decomposition.

The right panel of Fig.~\ref{fig:slopes_halos} reveals a qualitatively different behaviour for the BEC and mSFDM models. The BEC profile shows a rapid increase of the slope magnitude followed by sharp divergences at characteristic radii $r_0$. These features originate from the oscillatory form of the BEC density profile, where the density approaches zero at $r_0$, leading to formally divergent logarithmic slopes. The positions depend on the fitted halo scale radius and therefore differ slightly between the two baryonic mass models. The mSFDM profile exhibits a non-monotonic slope with alternating regions of steepening and flattening. This behaviour reflects the multistate structure of the SF halo, where the superposition of different quantum states produces radial oscillations in the density distribution. As a result, the density gradient may locally decrease or increase, which explains the sign changes of the slope.

\section{Discussion}\label{sec:Discussion}

The present analysis shows that the quality of the M31 RC reconstruction depends not only on the choice of the DM halo profile, but also on how realistically the baryonic structure of the galaxy is represented. In both the present SF analysis and the previously published study based on phenomenological halos, the two-bulge description leads to systematically lower BIC values than the single de Vaucouleurs bulge model. This agreement between two independent halo sectors indicates that the improvement is driven primarily by the internal structure of M31 rather than by a particular DM prescription. In other words, the observed kinematics in the central few kiloparsecs appear to require a more flexible baryonic decomposition, where a compact inner bulge and a more extended main bulge contribute separately to the gravitational potential.

Within this baryonic framework, the FDM model provides the best overall description of the M31 data among the SF profiles considered here. In the preferred two-bulge configuration, the FDM fit yields the minimum BIC of the present analysis, while the corresponding halo remains moderately compact and sufficiently massive to support the outer RC. By contrast, the BEC and mSFDM models are statistically disfavored, especially in the two-bulge case. This hierarchy is physically meaningful. The FDM profile is monotonic and cored, so it supplies a smooth transition from the baryon-dominated inner region to the halo-dominated outer region without introducing artificial radial features. Such behavior is well suited to the observed RC of M31, which exhibits a broad maximum and then a gradual decline at large radii.

The success of the FDM halo relative to the BEC and mSFDM cases suggests that, for M31, the data favor a DM distribution with a regular central core but without strong oscillatory structure. This conclusion is consistent with the logarithmic-slope analysis. The FDM slope remains smooth and slowly varying, which means that the density decreases in a controlled way and preserves the global stability of the halo contribution. In contrast, the BEC profile develops sharp slope divergences near the characteristic radius where the density tends to zero, while the mSFDM profile exhibits alternating steepening and flattening associated with the superposition of different states. Although such oscillatory behavior is mathematically admissible within SF scenarios, the present M31 data do not support it as the dominant halo pattern.

A comparison with the previous M31 study based on phenomenological halos \cite{2024IJMPD..3350016B} helps to clarify which conclusions are robust and which are model-dependent. In that work, the ES halo provided the best phenomenological fit, again in combination with the two-bulge baryonic model, while the present study finds that the best SF realization is the FDM halo. Therefore, the common feature of the preferred solutions is not the detailed functional form of the density profile, but the fact that both belong to the class of cored halos. This point is physically important. A cored halo contributes to the circular velocity in a gradual way and avoids the excessive central concentration characteristic of cuspy or strongly structured profiles. For M31, this appears to be crucial: the inner rotational support is already largely supplied by the stellar bulge and disk, so the halo must remain subdominant in the central region and become relevant only at intermediate and large radii. Both the ES halo in the phenomenological analysis and the FDM halo in the present work satisfy this requirement.

The fitted halo parameters also admit a straightforward interpretation. Relative to the BEC model, the preferred FDM halo is denser in the central region and characterized by a smaller scale radius, which means that it can provide the required mass more efficiently over the radial range probed by the data. The BEC halo, on the contrary, tends to become too extended and too diffuse, shifting a substantial fraction of its dynamical influence to radii where the observational constraints are weaker. This explains why the BEC fits can reproduce the broad trend of the RC but fail to match its detailed shape as successfully as FDM. The mSFDM case introduces additional internal structure through the coexistence of states, but the current RC data do not justify this extra complexity, as reflected by the larger BIC values. Thus, the statistical ranking of the models is directly connected to how effectively each profile distributes mass across the regions where M31 is actually observed.

Another outcome is that the inferred logarithmic slopes depend only weakly on whether the bulge is modeled as a single component or decomposed into inner and main bulges, at least for the smooth ES and FDM halos. This shows that once a given cored halo family is selected, its large-scale density gradient is determined mainly by the halo form itself rather than by moderate changes in the baryonic decomposition. Such stability is desirable from the physical point of view, because it means that the inferred halo structure is not an artifact of fine-tuning the visible components. By contrast, the more irregular behavior of the BEC and mSFDM slopes reflects genuine structural features of those models, not merely fitting noise. Hence the slope analysis supports the same conclusion reached from the BIC comparison: the M31 data favor a halo with a regular radial decline and disfavor profiles with strong oscillatory signatures.

\section{Conclusion} \label{sec:Conclusion}

In this work we investigated whether SFDM halo models can reproduce the observed RC of the M31 galaxy when the baryonic structure is modeled in a realistic way. To this end, we considered two alternative baryonic prescriptions, namely a single de Vaucouleurs bulge and a two-component bulge decomposition consisting of inner and main bulges described by ES profiles, together with a Freeman disk. Within this framework, we analyzed three SF motivated halo models, FDM, BEC and mSFDM, and compared their performance by means of nonlinear fitting and the Bayesian Information Criterion.

The results show that the adopted description of the baryonic sector plays an important role in the reconstruction of the M31 RC. In all cases considered in this work, the two-bulge configuration provides a better statistical description than the single-bulge one. This indicates that the inner kinematics of M31 are more naturally represented by a compact central structure plus a more extended main bulge, rather than by a single spheroidal component. Since the same conclusion was obtained in the earlier phenomenological analysis of M31 \cite{2024IJMPD..3350016B}, it should be regarded as a robust structural property of the galaxy rather than as a feature specific to a particular halo model.

Among the SF halos studied here, the FDM profile yields the best overall agreement with the data and gives the minimum BIC in the preferred two-bulge configuration. By contrast, the BEC and mSFDM models are less preferred. This shows that not all SF realizations are equally compatible with the M31 RC. The present data favor a halo with a smooth and monotonic cored density distribution, whereas profiles with stronger oscillatory structure are statistically disfavored.

A comparison with the previously published M31 study based on phenomenological halos suggests that the M31 RC primarily favors a halo with a moderate central concentration. Therefore, the main common result of the two studies is that the observational data are more consistent with cored halo configurations than with strongly cuspy or oscillatory ones.

At the same time, there are limitations of RC data as a unique discriminator of the DM nature. Although the FDM model slightly outperforms the best phenomenological reference halo considered here (ES), both descriptions reproduce the observed RC with comparable quality. Thus, the present analysis does not by itself establish the nature of the dark sector. Rather, it demonstrates that the FDM scenario is the most viable SF candidate among those examined in this work and remains competitive with the best phenomenological cored alternative.

Future studies could extend this analysis by applying the same modeling framework to a larger sample of galaxies in order to test the validity of the obtained results across different galactic environments. In addition, RC constraints can be combined with complementary observables such as stellar velocity dispersion, gravitational lensing or dynamical mass estimates.

\vspace{3mm}

\section{Acknowledgements}
\begin{acknowledgments}
This research was funded by the Science Committee of the Ministry of Science and Higher Education of the Republic of Kazakhstan (Grant No. AP23488743).
\end{acknowledgments}

\bibliography{0refs}
\end{document}